# Extended-p$^+$ Stepped Gate (ESG) LDMOS for Improved Performance

M. Jagadesh Kumar, *Senior Member, IEEE* and Radhakrishnan Sithanandam

*Abstract*—In this paper, we propose a new *E*xtended-p$^+$ *S*tepped *G*ate (ESG) thin film SOI LDMOS with an extended-p$^+$ region beneath the source and a stepped gate structure in the drift region of the LDMOS. The hole current generated due to impact ionization is now collected from an n$^+$p$^+$ junction instead of an n$^+$p junction thus delaying the parasitic BJT action. The stepped gate structure enhances RESURF in the drift region, and minimizes the gate-drain capacitance. Based on two-dimensional simulation results, we show that the ESG LDMOS exhibits approximately 63% improvement in breakdown voltage, 38% improvement in on-resistance, 11% improvement in peak transconductance, 18% improvement in switching speed and 63% reduction in gate-drain charge density compared with the conventional LDMOS with a field plate.

*Index Terms*—LDMOS, silicon on insulator (SOI), breakdown voltage, transconductance, on-resistance, gate charge

## I. INTRODUCTION

LATERALLY double diffused metal oxide semiconductor (LDMOS) on SOI substrate is a promising technology for RF power amplifiers and wireless applications [1-5]. In the recent past, developing high voltage thin film LDMOS has gained importance due to the possibility of its integration with low power CMOS devices and heterogeneous microsystems [6]. But realization of high voltage devices in thin film SOI is challenging because floating body effects affect the breakdown characteristics. Often, body contacts are included to remove the floating body effects in RF devices [7]. But with a body contact, the device area increases, and switching time and power density degrade. Another problem with the thin film SOI LDMOS is the kink appearing in the output characteristics. Though the kink can be eliminated by reducing the drift region doping concentration, the increased drift region resistance degrades the other device parameters [6]. Therefore, the motivation of this work is to explore structural modifications in SOI LDMOS to enhance its high voltage capabilities.

In this paper, we propose an *E*xtended-p$^+$ *S*tepped *G*ate LDMOS (ESG LDMOS) in which an extended-p$^+$ region beneath the source reduces the parasitic bipolar transistor effect improving the breakdown voltage and the stepped gate oxide improves RESURF and reduces the gate-drain charge density ($Q_{gd}$) leading to an improvement in the switching speed. Using two dimensional device simulations [8], the ESG LDMOS is shown to exhibit significant improvements in breakdown voltage, on-resistance, transconductance and gate-charge compared to the conventional LDMOS with a field plate. The extended-p$^+$ region, originally proposed in [9,10] to reduce the parasitic bipolar transistor effect in SOI MOSFETs, has recently been experimentally demonstrated in [11]. The stepped gate structure can be created using the conventional CMOS processing. We, therefore, believe that although our results are based on simulation, they may provide the incentive for experimental investigation of ESG LDMOS and its comparison with other structures.

Final manuscript received April 09, 2010. This work was supported in part by the IBM Faculty Award. The authors are with Department of Electrical Engineering, Indian Institute of Technology, New Delhi 110 016, India. (e-mail: mamidala@ieee.org).

## II. DEVICE STRUCTURE AND DESIGN ASPECTS

The schematic cross sections of the conventional LDMOS and the proposed ESG LDMOS are shown in Fig. 1. The conventional LDMOS structure used in our simulation is similar to the structure proposed by [6] and has a field plate over the drift region to enhance the high-voltage capabilities as shown in Fig. 1(a) [12]. However, the proposed device has an extended-$p^+$ region beneath the source and also a stepped gate oxide under the field plate as shown in Fig. 1(b). The main principle of the extended-$p^+$ region is to reduce the parasitic BJT effect [9-11]. The hole current generated due to impact ionization is now collected through an $n^+p^+$ junction instead of an $n^+p$ junction as shown in Fig. 2, and therefore the current gain of the parasitic BJT reduces (since the $p^+$ region now acts as the base). This low current-gain limits the collector current of the parasitic BJT and reduces the regenerative feedback resulting in higher breakdown voltage [10]. If the $p^+$ region is extended farther into the channel, the threshold voltage of the ESG LDMOS will increase. Therefore, we have taken the length of the $p^+$ region to be half the channel length to keep the threshold voltage of both the devices to be approximately the same. The stepped gate structure in the drift region has three gate sections arranged with increasing gate oxide thickness from the channel side to the drain side. The stepped gate structure leads to uniform electric field distribution, reduced gate-drain capacitance and facilitates the use of increased drift region doping simultaneously [13,14]. The gate overlap length and the stepped oxide thickness for both the devices in our simulation are designed for maximum breakdown voltage and minimum on-resistance. The parameters used in our simulations are given in Table. 1.

## III. PROPOSED PROCESSING STEPS

Fig. 3 shows the proposed fabrication procedure for the stepped gate of the ESG LDMOS. The processing steps are created using the process simulator ATHENA [15]. This is similar to the multi-gate formation on a stepped insulator reported by Xing et al. [16]. On an SOI wafer with an n-silicon layer, a 50 nm thick gate oxide is formed using thermal oxidation followed by a $p^+$ poly layer deposition and patterning. This allows the formation of a 1 μm long first gate on a 50 nm gate insulator (Fig. 3(a)). Subsequently, a 50 nm thick low temperature oxide (LTO) and over that a 0.4 μm thick $n^+$ polysilicon are deposited as shown in Fig. 3(b). Using a mask and photolithography, the polysilicon layer is etched leaving a sidewall polysilicon layer as shown in Fig. 3(c) which will now act as the second gate of 0.5 μm length on a 100 nm gate insulator. We now deposit a 200 nm LTO followed by a 0.4 μm thick $n^+$ polysilicon (Fig. 3(d)) and using a mask and photolithography, the polysilicon layer is etched resulting in the multigate structure shown in Fig. 3(e) with the third gate now on a 300 nm thick insulator. After depositing a thick conformal oxide, a chemical-mechanical polishing (CMP) process will planarize the gate as shown in Fig. 3(f). Once the gate is patterned, the rest of the fabrication is similar to that of any conventional LDMOS [11]. After the metallization process, the source, drain and the gate contacts are formed with all the three gates shorted resulting in the final ESG LDMOS structure shown in Fig. 1(b).

## IV. RESULTS AND DISCUSSION

We have created the conventional and the proposed ESG LDMOS using two dimensional device simulator ATLAS [8]. Appropriate models are invoked for impact ionization, SRH and Auger generation and recombination, carrier velocity saturation, concentration dependent mobility, transverse and vertical electric field dependent mobility [8].

### A. Drain Breakdown

The breakdown characteristics of the ESG LDMOS are compared with the conventional LDMOS as shown in Fig. 4. We have taken the breakdown voltage as the drain voltage at which the drain current exceeds 1 pA/$\mu$m at $V_{GS} = 0$ V. It can be seen that the breakdown voltage of the ESG LDMOS shows 63% improvement in the breakdown voltage. As compared to the conventional LDMOS with a field plate on a uniform gate oxide, the ESG LDMOS with a field plate on the stepped oxide exhibits a redistributed electric field across the drift region. From Fig. 5, we can see that for a given drain potential ($V_{DS} = 15$ V) and drift doping value, the stepped gate introduces additional electric field peaks which pulls down the main peak of the conventional LDMOS from 0.26 MV/cm to 0.12 MV/cm. With the enhanced source side depletion due to the stepped gate (and additional electric field peak formation), the drift region doping level of the ESG LDMOS can be increased with a simultaneous increase in breakdown voltage [17]. At higher $V_{DS}$, if the parasitic BJT turns on, the source-body (emitter-base) potential of the parasitic BJT should decrease. To examine the effect of the extended $p^+$ region on the parasitic BJT response, we have simulated the base-emitter (body-source potential) as a function of $V_{DS}$ for $V_{GS} = 0$ V and is shown in Fig. 6. It is clearly seen that for the ESG LDMOS, the body-source potential does not decrease until the drain voltage reaches about 48 V indicating the role of the extended $p^+$ region in suppressing the parasitic BJT.

### B. On-resistance

Fig. 7 shows the breakdown voltage as a function of drift region doping. It can be seen that the maximum breakdown of the conventional LDMOS with a field plate occurs when the drift region doping is around $4 \times 10^{16}$ cm$^{-3}$. Due to the stepped gate architecture [14], the optimal drift region doping for maximum breakdown voltage in the case of the proposed device is higher ($9 \times 10^{16}$ cm$^{-3}$) than the conventional LDMOS and therefore, will lead to a reduced on-resistance. The on-resistance is calculated as the ratio of applied drain voltage to the drain current when the device operates in linear region. Fig. 8 shows that the on-resistance of the ESG LDMOS improves by 38% over the conventional LDMOS.

### C. Transconductance

Transconductance has two attributes - peak transconductance (the maximum gate control that can be achieved) and range (the range of gate voltages for which the LDMOS responds). Both the peak transconductance and the gate voltage range should be higher for RF applications. As shown in Fig. 9, the peak transconductance of the proposed device shows an improvement of 11% compared to the conventional device and also the gate voltage range (i.e. range of gate voltage value, the device responds) extends to 6 V compared to 3 V of the conventional device resulting in 100% improvement. These improvements in peak transconductance and range are due to the reduced on-resistance of the ESG LDMOS. With reduced on-resistance, the quasi saturation or current compression is delayed (due to increased drift region doping) to higher gate potentials. Since the gate voltage range in which the device operates is increased due to the above mentioned reason, the peak transconductance shifts towards higher gate voltage values.

### D. Output Characteristics

Output characteristics determine the robustness of the LDMOS in RF amplification applications. Increased on-current also means a reduced device width for the given current specification. Therefore with ESG LDMOS, the chip area for the given output current decreases. As shown in Fig. 10, the ESG LDMOS exhibits improved current levels and no visible quasi-saturation region compared to the conventional device. The absence of quasi-saturation is a significant result.

### E. Gate Charging Transient

Gate charge, which gives an estimate of the switching losses incurred by the power device, is the amount of charge required on the gate to turn on the device for a given on-resistance. It should be as low as possible for the power device [18, 19]. Fig 11 shows the gate charge characteristics of the ESG LDMOS and the conventional device. The circuit used to calculate the gate charge in ATLAS mixed mode simulations is shown in the inset. The device width for both the structures is kept to be 10,000 $\mu$m. The input current source is chosen as 10 $\mu$A and the output current source is chosen as 100 $\mu$A. In Fig 11, the initial part of the curve till the first change in the slope gives the gate-source charge ($Q_{gs}$). The $Q_{gs}$ of the ESG LDMOS and the conventional LDMOS are obtained to be 72 pC/mm$^2$ and 33 pC/mm$^2$, respectively. The part of the curve with smaller slope gives the gate-drain charge ($Q_{gd}$) and is found to be 1006 pC/mm$^2$ and 377 pC/mm$^2$ for the conventional and the proposed device, respectively. Thus, we observe around 118% increase in $Q_{gs}$ and 63% decrease in $Q_{gd}$ for the ESG LDMOS as compared to the conventional LDMOS. At $V_{GS} = 4$ V, the total gate charge for the ESG LDMOS and the conventional LDMOS are 1083 pC/mm$^2$ and 2375 pC/mm$^2$, respectively, which is approximately 55% reduction in the gate charge.

### F. Switching Speed

Switching characteristics play an important role in high frequency operations of the power device. For calculating the switching delay, we have constructed the inverter circuit shown in the inset of Fig. 12 using ATLAS mixed mode option. The device width is chosen to be 45 $\mu$m for both the conventional and the proposed device. We have applied a 5 V input pulse with 50 ps rise time to the input terminal to observe the output. The switching delay is calculated as the time difference between the input and output signal at 2.5 V (50% of applied voltage). It can be seen from Fig. 12 that the switching delay of the ESG LDMOS is 28.2 ps compared to 33.2 ps for the conventional LDMOS. This is approximately 18% reduction in switching delay. This improvement is primarily due to the reduced on-resistance and gate-to-drain capacitance ($C_{gd}$) in the proposed device (as discussed in the above section).

## V. CONCLUSIONS

In this work, we have proposed a new ESG LDMOS in which an extended-p$^+$ region beneath the source, eliminates the parasitic bipolar transistor effect and the stepped gate structure enhances the gate-charge behavior. By combining these structural changes, significant improvements in LDMOS device characteristics are reported. When compared to a conventional LDMOS with a field plate, the proposed ESG LDMOS is shown to exhibit 63% increase in breakdown voltage, 38% reduction in on-resistance, 11% increase in peak transconductance, 118% increase in $Q_{gs}$ and 63% decrease in $Q_{gd}$ and 18% improvement in switching speed.

**Table 1**

**CONVENTIONAL AND ESG LDMOS DEVICE PARAMETERS**

**IN DEVICE SIMULATIONS**

| | |
|---|---|
| Gate oxide thickness, ($t_{ox1}$, $t_{ox2}$ and $t_{ox3}$) | 50 nm, 100 nm and 300 nm |
| Gate length, ($L_{G1}$, $L_{G2}$ and $L_{G3}$) | 1 μm, 0.5 μm and 0.5 μm |
| Oxide spacers between stepped gates ($L_{SP1}$, $L_{SP2}$) | 50 nm, 200 nm |
| Channel length, L | 0.5 μm |
| Drift region length | 3.5 μm |
| Silicon film thickness | 200 nm |
| Buried oxide thickness | 400 nm |
| Source/Drain doping | $1 \times 10^{19}$ cm$^{-3}$ |
| P-body doping | $1 \times 10^{17}$ cm$^{-3}$ |
| Drift region doping (conventional & ESG LDMOS) | $4 \times 10^{16}$ cm$^{-3}$, $9 \times 10^{16}$ cm$^{-3}$ |
| Extended-p$^+$ doping | $1 \times 10^{19}$ cm$^{-3}$ |
| Threshold voltage (conventional & ESG LDMOS) | ≈1.85 V |

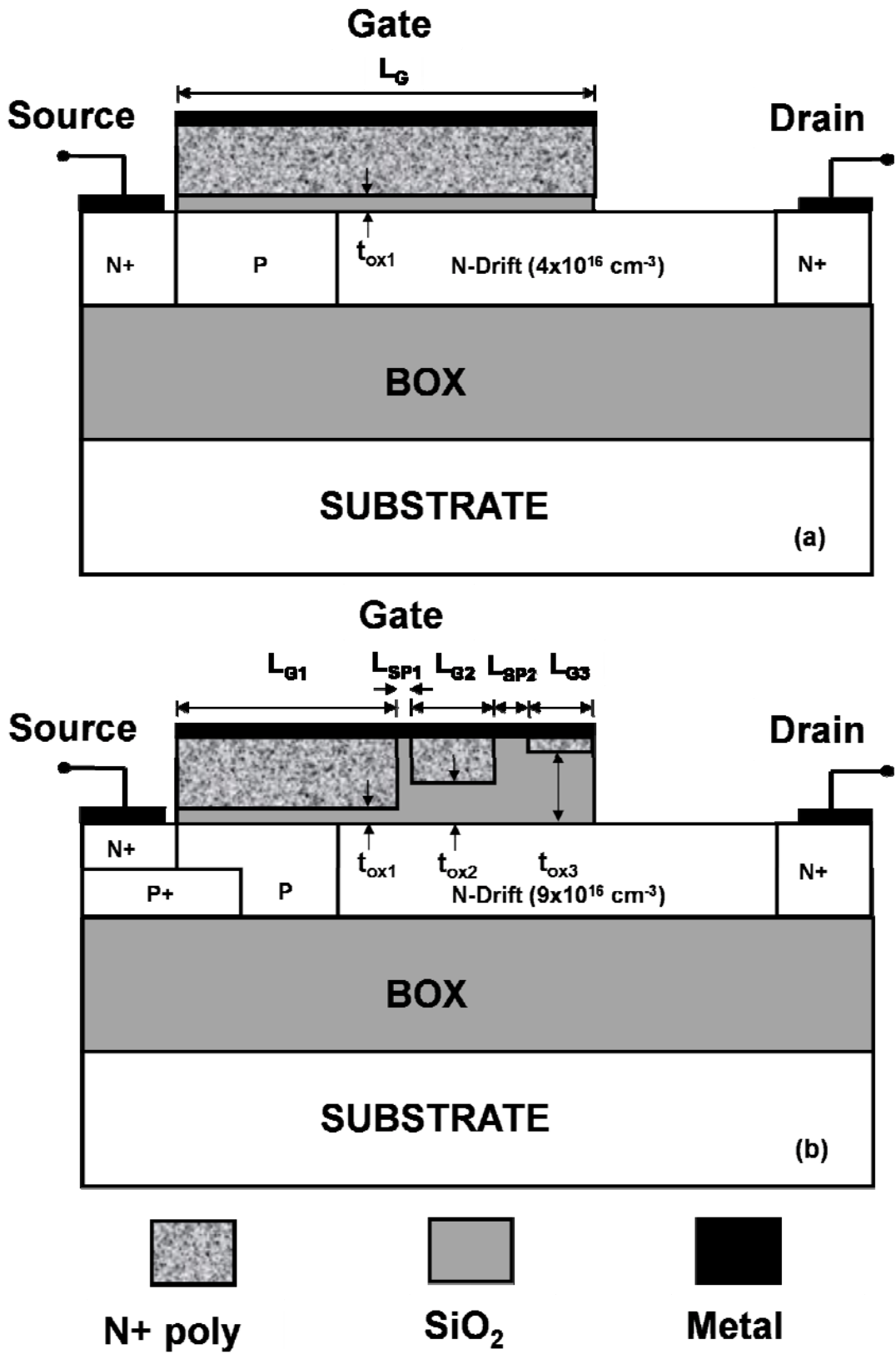

Fig. 1. (a) Cross-sectional view of conventional LDMOS (b) ESG LDMOS.

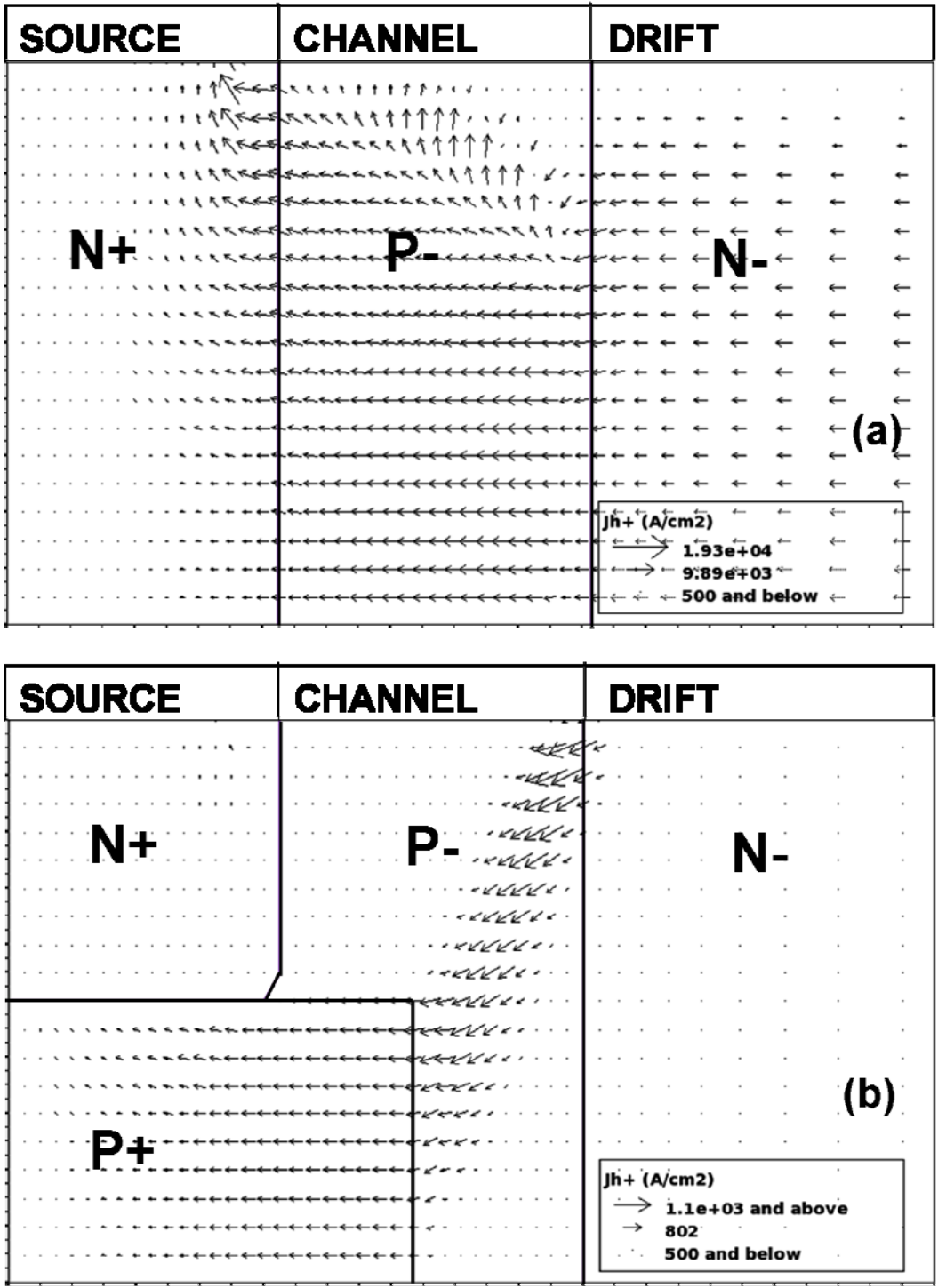

Fig. 2. Hole current vectors across source and channel of (a) conventional LDMOS (b) ESG LDMOS at $V_{GS} = 4$ V and $V_{DS} = 20$ V.

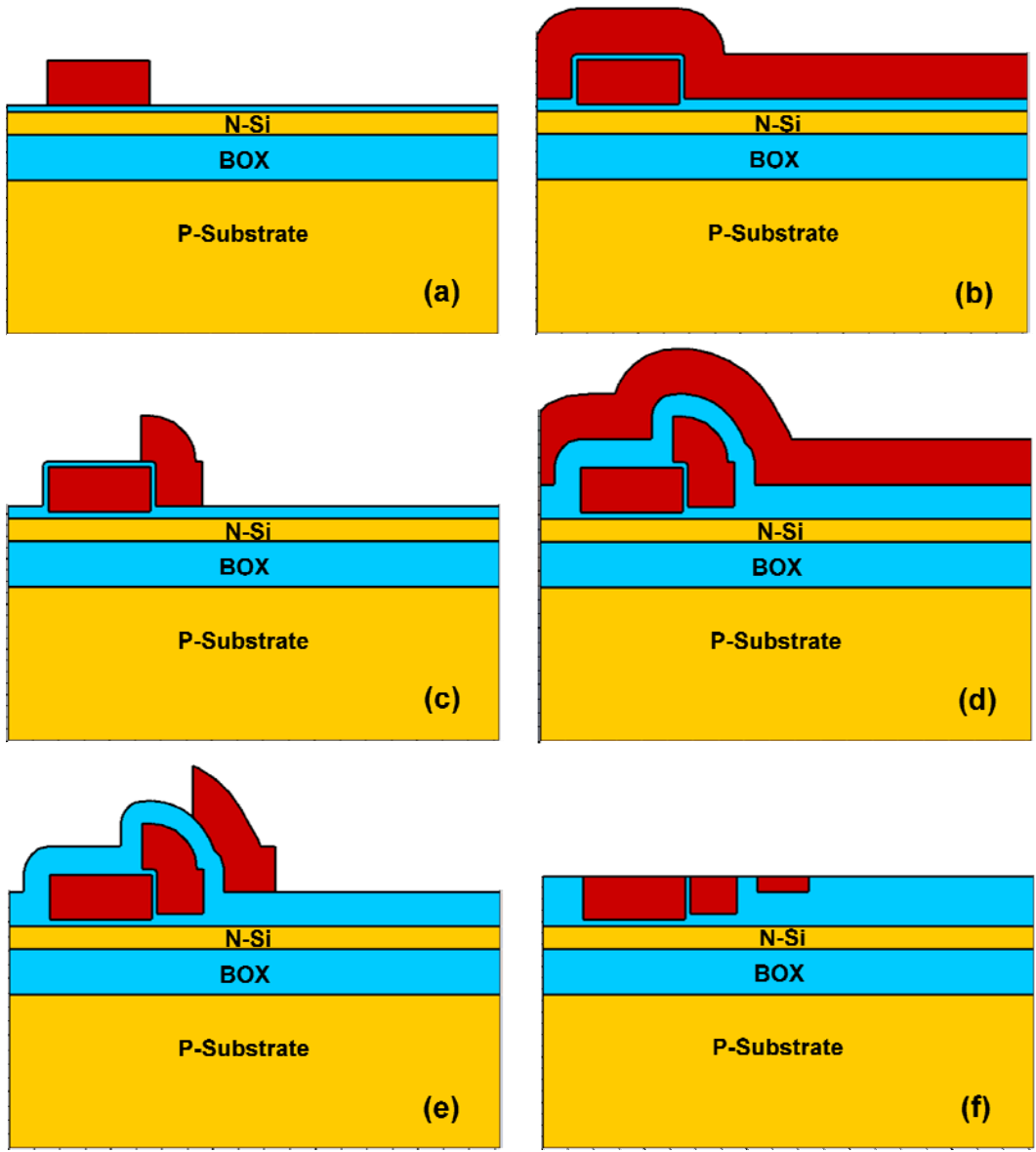

Fig. 3. ATHENA generated process steps for forming the multi-gate structure on a stepped insulator for the ESG LDMOS.

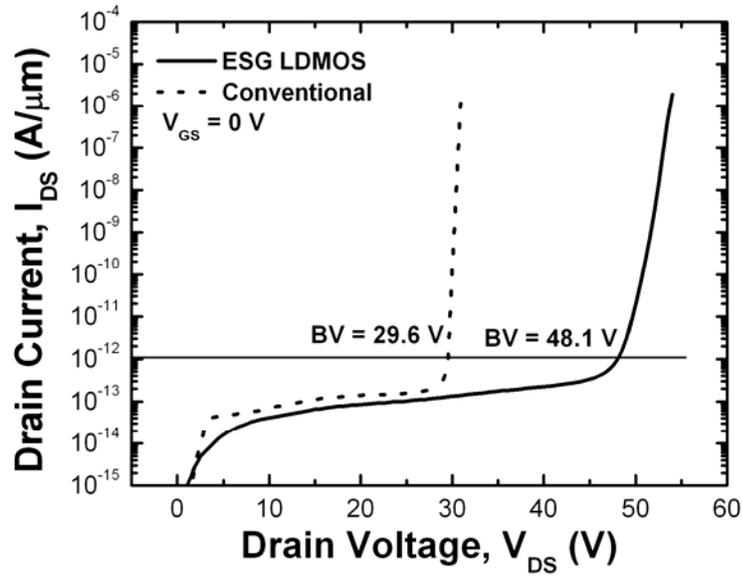

Fig. 4 Breakdown voltage characteristics of the ESG and the conventional LDMOS.

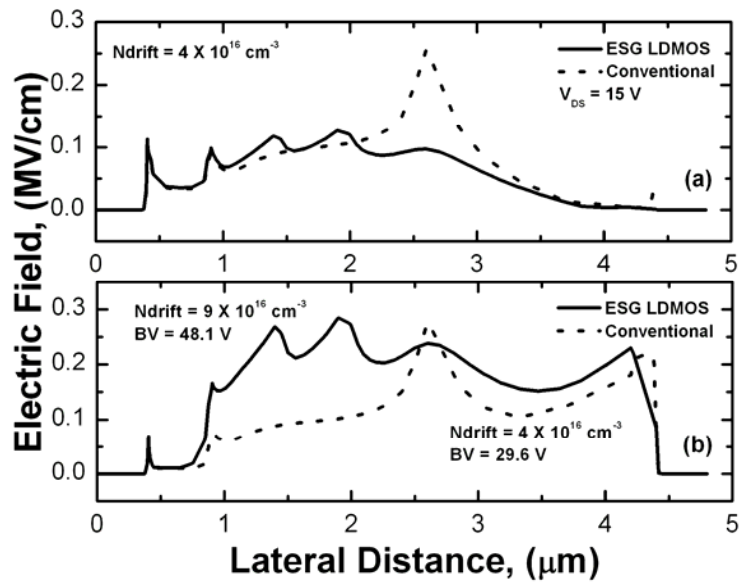

Fig. 5  Electric field distribution along the surface of the ESG and the conventional LDMOS (a) at $V_{DS}$ = 15 V and with same drift doping level (b) at breakdown voltage and with optimum drift region doping level.

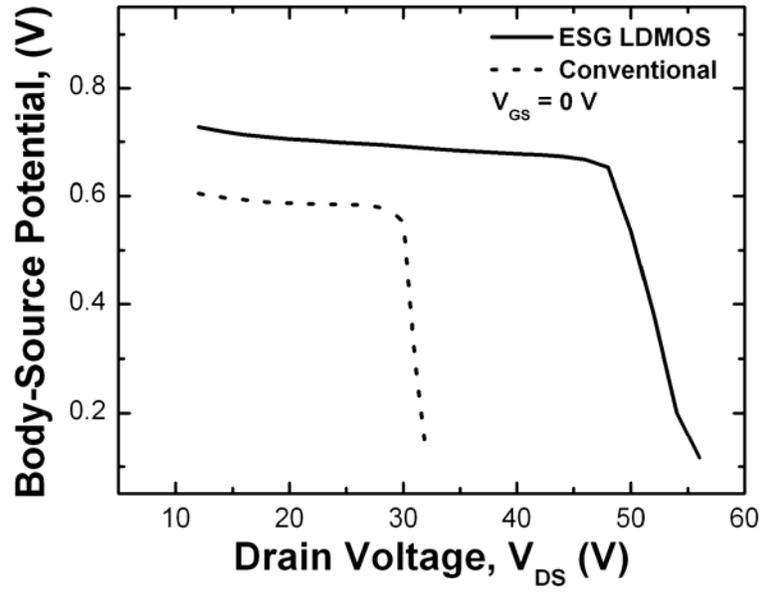

Fig. 6 Base emitter (body-source) internal potential variation with varying drain voltages of the ESG and the conventional LDMOS.

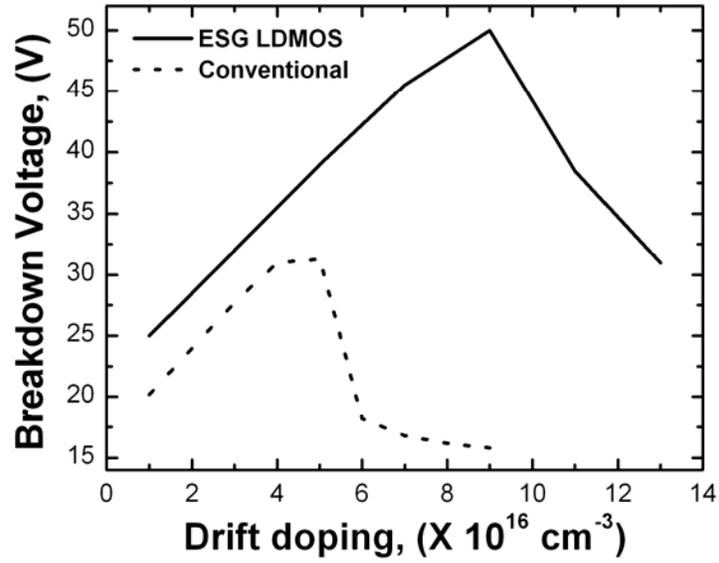

Fig.7 Breakdown voltage as a function of drift region doping of the ESG and the conventional LDMOS.

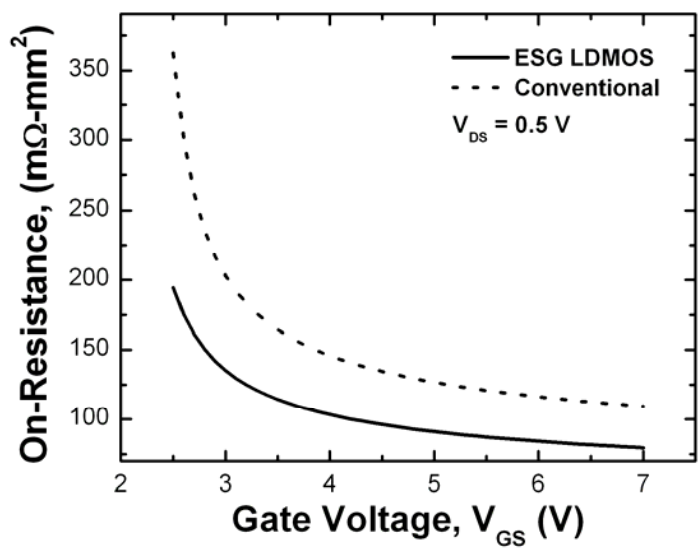

Fig. 8 On-resistance of the ESG and the conventional LDMOS.

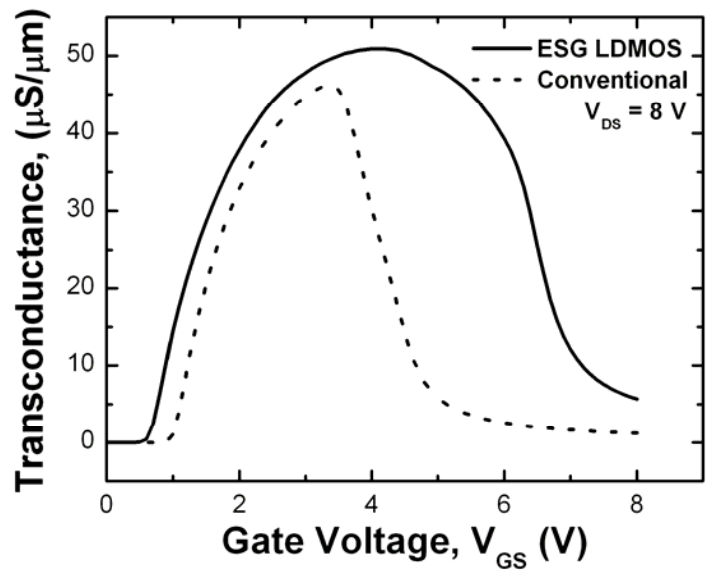

Fig. 9 Transconductance variation of the ESG and the conventional LDMOS.

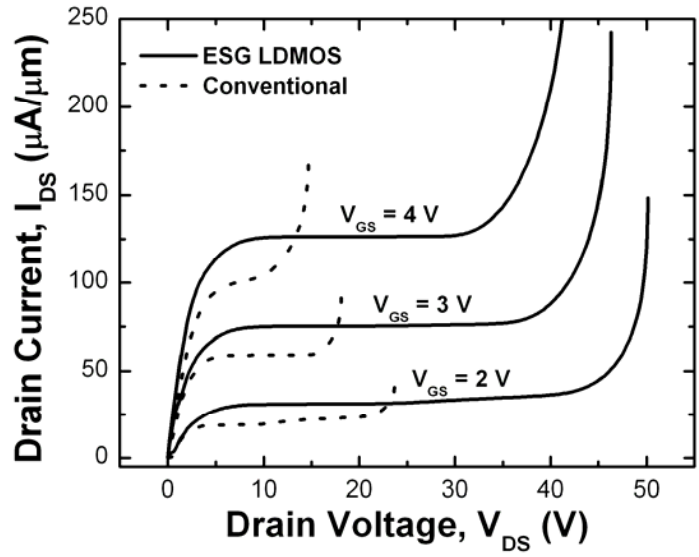

Fig. 10 Output characteristics of the ESG and the conventional LDMOS.

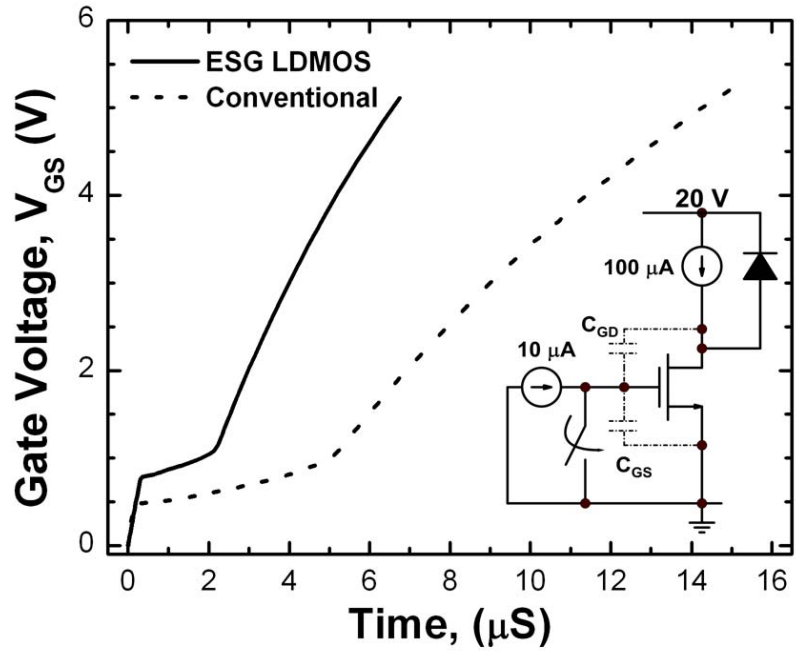

Fig. 11 Gate charging curve of the ESG and the conventional LDMOS.

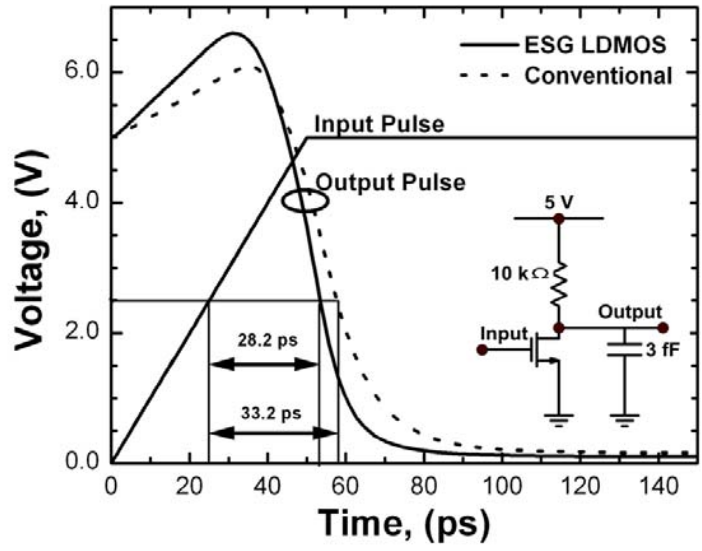

Fig. 12 Switching characteristics of the ESG and the conventional LDMOS.

**M. Jagadesh Kumar** was born in Mamidala, Andhra Pradesh, India. He received the M.S. and Ph.D. degrees in electrical engineering from the Indian Institute of Technology (IIT), Madras, India. From 1991 to 1994, he performed a postdoctoral research in the modeling and processing of high-speed bipolar transistors with the Department of Electrical and Computer Engineering, University of Waterloo, Waterloo, ON, Canada, where he also did a research on amorphous silicon TFTs. From July 1994 to December 1995, he was initially with the Department of Electronics and Electrical Communication Engineering, IIT, Kharagpur, India, and then, he joined the Department of Electrical Engineering, IIT, Delhi, India, where he became an Associate Professor in July 1997 and has been a Full Professor since January 2005. He is the Coordinator of the *VLSI Design, Tools and Technology* interdisciplinary program at IIT Delhi. His research interests include nanoelectronic devices, device modeling and simulation for nanoscale applications, integrated-circuit technology, and power semiconductor devices. He has published extensively in these areas of research with three book chapters and more than 145 publications in refereed journals and conferences. His teaching has often been rated as outstanding by the Faculty Appraisal Committee, IIT Delhi.

Dr. Kumar is a Fellow of the Indian National Academy of Engineering and the Institution of Electronics and Telecommunication Engineers (IETE), India. He is an IEEE Distinguished Lecturer of the Electron Devices Society. He is a member of the EDS Publications Committee and the EDS Educational Activities Committee. He is an Editor of the IEEE TRANSACTIONS ON ELECTRON DEVICES. He was the lead Guest Editor for (i) the joint special issue of the IEEE TRANSACTIONS ON ELECTRON DEVICES and IEEE TRANSACTIONS ON NANOTECHNOLOGY (November 2008 issue) on Nanowire Transistors: Modeling, Device Design, and Technology and (ii) the special issue of the IEEE TRANSACTIONS ON ELECTRON DEVICES on Light Emitting Diodes (January 2010 issue). He is the Editor-in-Chief of the *IETE Technical Review* and an Associate Editor of the *Journal of Computational Electronics.* He is also on the editorial board of *Recent Patents on Nanotechnology, Recent Patents on Electrical Engineering, Journal of Low Power Electronics,* and *Journal of Nanoscience and Nanotechnology*. He has reviewed extensively for different international journals.

He was a recipient of the 29th IETE Ram Lal Wadhwa Gold Medal for his distinguished contribution in the field of semiconductor device design and modeling. He was also the first recipient of the *ISA-VSI TechnoMentor Award* given by the India Semiconductor Association to recognize a distinguished Indian academician for playing a significant role as a Mentor and Researcher. He is also a recipient of the 2008 IBM Faculty Award. He was the Chairman of the Fellowship Committee of The Sixteenth International Conference on VLSI Design (January 4–8, 2003, New Delhi, India), the Chairman of the Technical Committee for High Frequency Devices of the International Workshop on the Physics of Semiconductor Devices (December 13–17, 2005, New Delhi), the Student Track Chairman of the 22nd International Conference on VLSI Design (January 5–9, 2009, New Delhi), and the Program Committee Chairman of the Second International Workshop on Electron Devices and Semiconductor Technology (June 1–2, 2009, Mumbai, India).

**S. Radhakrishnan** received the B.Tech degree in Electronics and Communication from Pondicherry Engineering College, Pondicherry, India in 2005. He is currently working towards the MS degree by Research from Department of Electrical Engineering at Indian Institute of Technology Delhi, India. His research interests includes power semiconductor devices, nanoscale devices and modeling. He has published about 5 papers in various international refereed journals and conference proceedings.